\documentclass[prbpreprint,showpacs,preprintnumbers,amsmath,amssymb,superscriptaddress]{revtex4}

\usepackage{graphicx}
\usepackage{amssymb}

\begin{document}

\title{Theory of itinerant magnetic excitations in the SDW phase of iron-based superconductors}

\author{J.~Knolle}
\affiliation {Max-Planck-Institut f\"{u}r Physik komplexer Systeme, D-01187 Dresden, Germany}

\author {I.~Eremin}
\affiliation {Max-Planck-Institut f\"{u}r Physik komplexer Systeme, D-01187 Dresden, Germany}
\affiliation {Institut f\"{u}r Mathematische und Theoretische Physik, TU Braunschweig, D-38106
Braunschweig, Germany}

\author{A.V.~Chubukov}
 \affiliation{Department of Physics, University of Wisconsin-Madison, Madison,
Wisconsin 53706, USA}

\author{R.~Moessner}
\affiliation {Max-Planck-Institut f\"{u}r Physik komplexer Systeme, D-01187 Dresden, Germany}

\begin{abstract}
We argue that salient experimental features of the magnetic excitations in the SDW phase of iron-based superconductors can be understood within an itinerant model. We identify a minimal model and use a multi-band RPA treatment of the dynamical spin susceptibility.  Weakly-damped spin-waves are found near the ordering momentum and it is shown how they dissolve into the particle-hole continuum. We show that ellipticity of the electron bands accounts for the anisotropy of the spin waves along different crystallographic directions and the gap at the momentum conjugated to the ordering one.
We argue that our theory agrees well with the neutron scattering data.
\end{abstract}
\pacs{74.70.Xa,75.10.Lp,75.30.Ds}
\date{\today}

\maketitle

{\it Introduction.} The newly discovered iron arsenide superconductors\cite{kamihara} opened a new field in the research on high $T_c$ superconductivity. The phase diagram of ferro-pnictide (FP) superconductors is similar to that of layered cuprates and contains an antiferromagnetic phase at small dopings and a
superconducting phase at larger dopings. However, parent compounds of iron-based systems are antiferromagnetic metals rather than Mott insulators, in contrast to the cuprates.
The electronic structure of parent compounds in the normal state
consists of two circular hole pockets of unequal size, centered around the $\Gamma$ point $(0,0)$,  and two elliptic electron pockets
centered at $(0, \pm \pi)$ and $(\pm \pi,0)$  points in the
unfolded Brillouin zone (UBZ), which is solely based on the $Fe$-lattice~\cite{LDA,kaminski,coldea}. The  dispersions of electron and hole bands are significantly nested, {\it i.e.} $\varepsilon_{\bf k}^{h} \simeq -\varepsilon_{\bf k+Q_i}^{e}$ where ${\bf Q}_i$ is either ${\bf Q}_1 = (\pi,0)$ or ${\bf Q}_2=(0,\pi)$.
Such nesting is a boost for antiferromagnetism, and several researchers argued that magnetism is itinerant, and at least partly comes from nesting~\cite{Tesanovic,eremin,brydon_new}. Others argued that magnetism is almost localized and is best described by the $J_1-J_2$ model~\cite{zhao}.

Neutron scattering  measurements on parent $FeAs$ compounds revealed that magnetic order  consists of ferromagnetic chains along one crystallographic direction in an $Fe$ plane and antiferromagnetic chains along the other~\cite{cruz,klauss}, i.e., the system selects   ${\bf Q}_i$ to be either  ${\bf Q}_1$ {\it or}
 ${\bf Q}_2$. Sharp propagating spin-wave like excitations have been observed near the ordering momentum (e.g.,  ${\bf Q}_1$) up to energies of around 100 meV, with different velocities along the two crystallographic directions~\cite{zhao,Diallo}. At larger energies, excitations become overdamped ~\cite{Diallo}.
Such  stripe order arises in the $J_1-J_2$ model of localized spins
for $J_2 < 0.5 J_1$, and it was argued that such a model can successfully describe some of the experimental data on the spin wave spectra, although, an unusually large in-plane anisotropy of the antiferromagnetic exchange between nearest neighbor spins has to be assumed\cite{zhao}. However, the same stripe order
 also appears in the itinerant model as a spin-density-wave (SDW) state with
 either  ${\bf Q}_1$ or
 ${\bf Q}_2$,
and is stabilized by the ellipticity of the electron bands and the interactions between the two electron pockets ~\cite{eremin}. Furthermore, within the itinerant model, only one hole and one electron Fermi surface (FS) is involved in the SDW mixing. The other two FSs remain intact and give rise to metallic behavior in the SDW phase, even at perfect nesting.

The issue we address here is whether neutron measurements of magnetic excitations can be explained within the itinerant model.
This is a crucial test of
the itinerant description of magnetism in FPs.
A first step in this direction was made in Ref.~\cite{brydon_new}, who analyzed  spin excitations in the SDW phase of FPs within a two-band model with circular hole and electron pockets.
We argue  that to describe anisotropic magnetic excitations in the SDW state one has to consider the
model consisting of one circular hole pocket centered around the $\Gamma$-point and two elliptic electron pockets centered around the $(\pi,0)$ and $(0,\pi)$ points of the unfolded BZ, respectively.
Based on this model, we develop a multi-band RPA treatment of the
dynamical spin susceptibility and compare the results with available experimental data.  We argue that both the high-energy particle-hole continuum {\it and} the low-energy propagating excitations can be quantitatively reproduced within our itinerant model.  We show that the fact that only one out of two electron pockets is involved in the SDW is responsible for the observed anisotropy of the spin wave excitations along different crystallographic  directions.

\begin{figure}[t]
\centering
\includegraphics[width=1.0\linewidth]{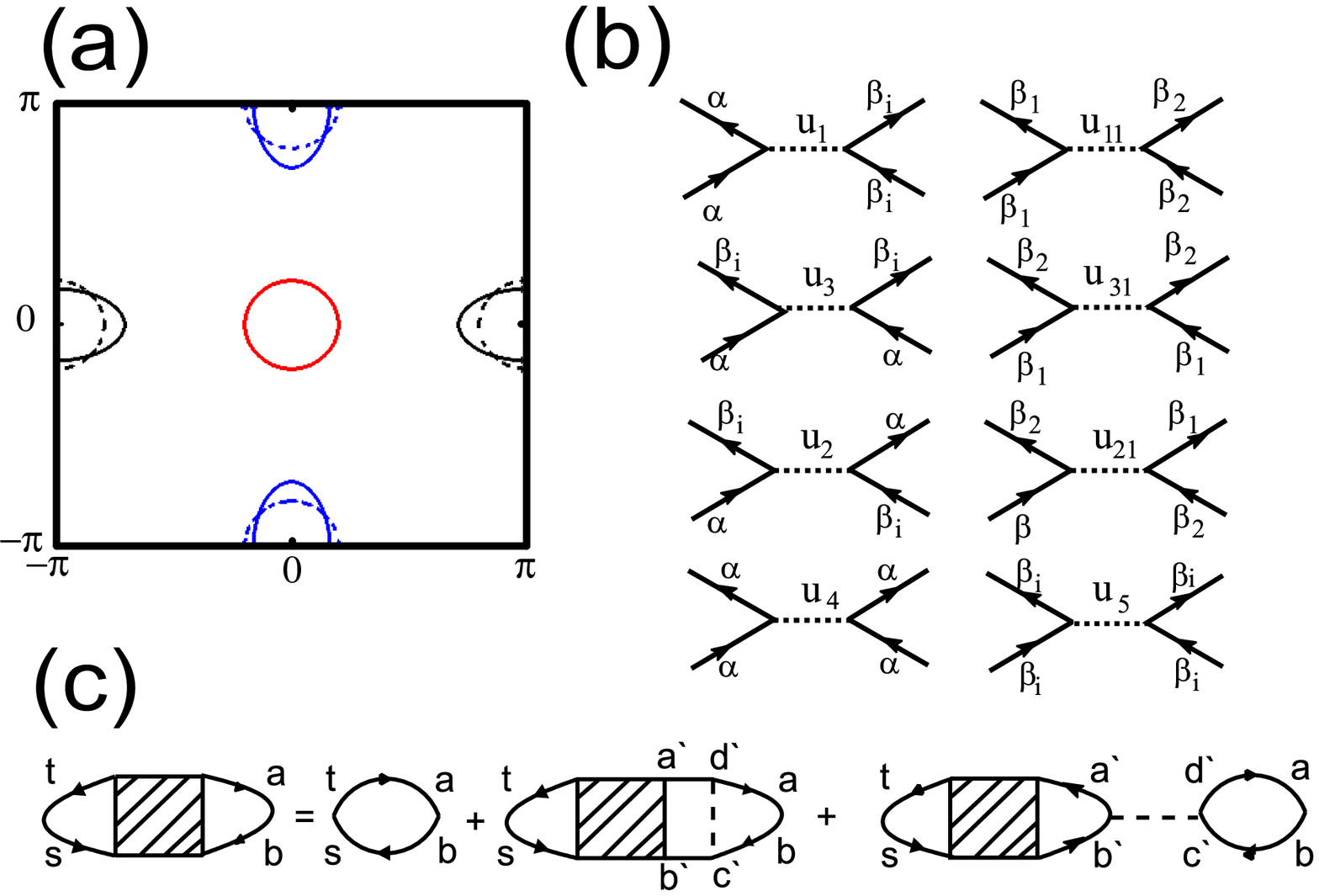}
\caption{(color online) (a) Calculated Fermi surfaces for the three band model. The dashed curves refer to the case of $\epsilon=0$, {\it i.e.},
complete nesting; Diagrammatic representations of the density-density type interactions (b) and the Dyson equation for the RPA spin susceptibility (c) in the three band model.}
\label{fig1}
\end{figure}

{\it The model.} We depart from a 4-band model, use the experimental fact that the two hole FSs  around the $\Gamma$-point have quite different
sizes and assume that one hole band interacts with the electron bands
 much weaker than the other and is thus not important for magnetism.
This leads to an effective three band model with one circular hole
pocket at $(0,0)$  ($\alpha$-band) and two elliptic electron pockets at
${\bf Q}_1$ and ${\bf Q}_2$ ($\beta$-bands):
\begin{eqnarray}
\label{eqH}
\lefteqn{H_2  =} && \nonumber\\
&& \sum_{\mathbf{p}, \sigma} \left[ \varepsilon^{\alpha}_{\mathbf{p}} \alpha_{\mathbf{p}  \sigma}^\dag \alpha_{\mathbf{p} \sigma}
+ \varepsilon^{\beta_{1}}_{\mathbf{p}} \beta_{1\mathbf{p} \sigma}^\dag \beta_{1\mathbf{p} \sigma} + \varepsilon^{\beta_{2}}_{\mathbf{p}} \beta_{2 \mathbf{p} \sigma}^\dag \beta_{2\mathbf{p} \sigma} \right].
\end{eqnarray}
We consider lattice dispersions for all three bands and set
 $\varepsilon^{\alpha}_{\mathbf{p}} = t_\alpha\left( \cos p_x +\cos p_y \right) -\mu$ and
$\varepsilon^{\beta_1}_{\mathbf{p}}= \epsilon_0 + t_\beta\left( \left[1+\epsilon \right]\cos(p_x+\pi)+\left[ 1-\epsilon \right]\cos(p_y)\right) -\mu$,
 $\varepsilon^{\beta_2}_{\mathbf{p}}= \epsilon_0 +
t_\beta\left( \left[1-\epsilon \right]\cos(p_x)+\left[ 1+\epsilon \right]\cos(p_y+\pi) \right)-\mu$. The parameter $\epsilon$ accounts for the ellipticity of the electron pockets. To make qualitative as well as quantitative comparisons to experiments, we use the Fermi velocities and the size of the Fermi pockets based on Refs. \cite{LDA,Ding}. We obtain $t_\alpha=0.85 eV$, $t_\beta=-0.68 eV$,
 $\mu=1.54 eV$, $\epsilon_0 = 0.31 eV$, and $\epsilon=0.5$.
For these values, Fermi velocities are $0.5eV a$ for the $\alpha$-band, where $a$ is the $Fe-Fe$ lattice spacing,
 and $v_x=0.27 eV a$ and $v_y=0.49 eV a$ along  $x$- and $y$-directions for
 the $\beta_1$-band (vice versa for $\beta_2$, we set $a_x = a_y =a$).
The corresponding Fermi surfaces is shown in Fig. \ref{fig1}(a).

The interacting part of the Hamiltonian contains
density-density interactions with small momentum transfer and momentum transfers $(\pi,0)$, $(0,\pi)$, and $(\pi,\pi)$. The interactions contributing to SDW formation are (in the terminology of Ref.\cite{Chubukov08})
\begin{eqnarray}
&&H_{4}= \sum  U_1 {\alpha}^{\dagger}_{{\bf p}_3 \sigma}
{\beta}^{\dagger}_{j{\bf p}_4 \sigma'}  {\beta}_{j{\bf p}_2 \sigma'} {\alpha}_{{\bf p}_1
\sigma}  + \nonumber \\
&& \frac{U_3}{2}
\left[{\beta}^{\dagger}_{j{\bf p}_3 \sigma} {\beta}^{\dagger}_{j{\bf p}_4 \sigma'}
{\alpha}_{{\bf p}_2 \sigma'} {\alpha}_{{\bf p}_1 \sigma} + h.c \right].
 \label{eq:2}
\end{eqnarray}
For simplicity, we treat both $U_1$ and $U_3$ as constants.
Both interactions involve
$\beta_1$ and $\beta_2$ fermions and in general give rise to two SDW OPs
${\vec \Delta}_1 \propto \sum_{\bf p} \langle \alpha^\dag_{\mathbf{p}\delta} \beta_{1\mathbf{p+Q_1} \gamma}
\vec{\sigma}_{\delta \gamma} \rangle$  with momentum  ${\bf Q}_1 = (\pi,0)$ and ${\vec \Delta}_2 \propto \sum_{\bf p} \langle \alpha^\dag_{\mathbf{p}\delta} \beta_{2\mathbf{p+Q_2} \gamma}  \vec{\sigma}_{\delta \gamma} \rangle$
with momentum  ${\bf Q}_2 = (0,\pi)$.  For perfect nesting and when $\beta_1$
 and $\beta_2$ fermions do not interact directly, the energy depends only on
${\vec \Delta}^2_1 + {\vec \Delta}^2_2$, i.e., the order is in general a combination of ${\vec \Delta}_1$ and ${\vec \Delta}_2$ and the ground state is degenerate. However, ellipticity of the electron pockets and a direct interaction between electron bands gives rise to an additional $a_1 {\vec \Delta}_1^2 {\vec \Delta}_2^2$ term in the energy, with a positive prefactor $a_1$ (Ref. \cite{eremin}). Then, the energy is minimized when either ${\vec \Delta}_1 =0$ or ${\vec \Delta}_2 =0$, i.e., SDW order is ferromagnetic along one direction and antiferromagnetic along the other, in agreement with experiments. Such a state couples $\alpha$ fermions with only one band of $\beta$ fermions, leaving the other band intact, i.e., leaving one of the electron FSs unaffected by SDW.

Without loss of generality we direct $\vec{\Delta}_1$ along the $z$-axis
and set $\vec{\Delta}_2 =0$.
The resulting mean-field Hamiltonian
can be diagonalized by a standard Bogolyubov transformation
\begin{eqnarray}
\alpha_{{\bf p}\sigma} & = & u_{{\bf p}} c_{{\bf p}\sigma}+v_{{\bf p}} d_{{\bf p}\sigma}, \nonumber \\
\beta_{1 {\bf p+Q_1}\sigma} & = & \sigma(v_{{\bf p}} c_{{\bf p}\sigma}-u_{{\bf p}} d_{{\bf p}\sigma}).
\end{eqnarray}
where $u_{{\bf p}}^2 \, , v_{{\bf p}}^2  = \frac{1}{2}
\left( 1\pm \frac{\varepsilon^{\alpha}_{\bf{p}}-\varepsilon^{\beta_1}_{\bf{p+Q_1}}}{\sqrt{
(\varepsilon^{\alpha}_{\bf{p}}-\varepsilon^{\beta_1}_{\bf{p+Q_1}})^2+4\Delta_1^2}}\right)$.
After the transformation, the Hamiltonian $H_2 + H_4$ becomes
\begin{equation}
H_{SDW}^{diag}=  \sum_{{\bf p},\sigma} \left[  E_{{\bf p}}^c c_{{\bf p}\sigma}^{\dagger}
c_{{\bf p}\sigma}+E_{{\bf p}}^d d_{{\bf p}\sigma}^{\dagger} d_{{\bf p} \sigma}+
\varepsilon^{\beta_2}_{\bf{p}} \beta_{2 \bf{p} \sigma}^\dag \beta_{2\bf{p} \sigma} \right],
\end{equation}
where
\begin{equation}
E_{{\bf p}}^{c,d}=
\frac{1}{2}\left( \varepsilon^{\alpha}_{\bf{p}}+\varepsilon^{\beta_1}_{\bf{p+Q_1}}
\pm \sqrt{(\varepsilon^{\alpha}_{\bf{p}}-\varepsilon^{\beta_1}_{\bf{p+Q_1}})^2+4\Delta_1^2} \right)
\end{equation}
 are the new dispersions and
the magnitude of the SDW gap is determined self-consistently from
$ \Delta_1 = -U_{SDW} \sum_{\bf p} u_{{\bf p}} v_{{\bf p}} \left[ f(E_{{\bf p}}^c)-f(E_{{\bf p}}^d) \right]$, where $U_{SDW} =U_1+U_3$ and $f(E)$ is the Fermi function.

{\it Magnetic Susceptibility and RPA.}
We will follow earlier works~\cite{brydon_new,Graser} and use a
 generalized RPA approach.
The dynamical susceptibility tensor in the multiband case is defined as
\begin{eqnarray}
\chi^{st, ab}_{lm}({\bf q},{\bf q}', i\Omega) = \frac{1}{2} \int_{0}^{\beta} d\tau e^{i \Omega \tau} \sum_{{\bf p},{\bf p'},
\gamma, \delta, \gamma', \delta'}\times \nonumber\\
\langle T_{\tau} s_{{\bf p} \gamma}^{\dagger} (\tau) t_{{\bf p+q} \delta}
(\tau) a_{{\bf p'} \gamma'}^{\dagger} (0) b_{{\bf p'-q'} \delta'}(0) \rangle
\sigma^l_{\gamma \delta} \sigma^m_{\gamma' \delta'} \quad,
\end{eqnarray}
where $T_{\tau}$ is the time ordering operator and quasiparticle operators
$s, t, a, b$ are operators from either $\alpha$ or $\beta_{1,2}$ bands.
Because  the unit cell is doubled in the SDW state, the
susceptibility is nonzero for {\bf q=q$'$} and {\bf q=q$'$+Q}$_1$~(Ref. \cite{Schrieffer}). In terms of Green functions (GF) we have
$\chi^{st,ab}_{zz}({\bf q},i\Omega)=-\frac{1}{2\beta} \sum_{\omega_n}
\sum_{{\bf p}, \sigma} G^{bs}_{{\bf p} \sigma}(i\omega_n) G^{ta}_{{\bf p+q} \sigma} (i\omega_n +i\Omega)$,
and
$\chi^{st,ab}_{\pm}(q,i\Omega)=-\frac{1}{\beta} \sum_{\omega_n} \sum_{{\bf p}} G^{bs}_{{\bf p} \uparrow}
(i\omega_n) G^{ta}_{{\bf p+q} \downarrow} (i\omega_n +i\Omega)$,
 where $G_{{\bf p} \sigma}^{st}(i\omega_n)=-\int_{0}^{\beta} d \tau \langle T_{\tau}
s_{{\bf p}\sigma}(\tau)t_{{\bf p}\sigma}^{\dagger}(0)
\rangle e^{i w_n \tau}$.
In order to compute the RPA susceptibility,
we also include all other interactions: the exchange interaction between $\alpha$ and $\beta$ fermions and the interactions between electron pockets (see Fig.
\ref{fig1}(b)).  These interactions do not contribute towards SDW formation but they do play a role in determining the structure of the spin wave excitations.

Spin susceptibilities $\left[\chi^{st,ab}_{lm}\right]_{RPA}$ in the RPA approximation
 are obtained via a Dyson equation
\begin{equation}
\left[\chi^{st,ab}_{lm}\right]_{RPA}=\chi^{st,ab}_{lm}+\chi^{st,a'b'}_{lm}
 U^{a'b',s't'}_{lm}\chi^{s't',ab}_{lm} \quad,
\label{n_1}
\end{equation}
and the summation over repeated band indices is assumed. The
Dyson equation is schematically shown in Fig.\ref{fig1}(c) (depending on
$l,m$, the series contain either ladder or bubbles).
The solution of Eq.(\ref{n_1}) in matrix form is straightforward
$\left[\hat\chi_{lm}\right]_{RPA}=\hat\chi_{lm}(1-\hat U_{lm}\hat\chi_{lm})^{-1}$.
For the single band Hubbard model our results agree with those of Ref.\cite{Schrieffer}.

{\it Results and Comparison to Experiments.}
We set $U_{SDW} \approx 0.52$eV for zero ellipticity, to match
the experimental $T_N \sim 200K$. This yields a $T=0$ value of the SDW gap of
$\Delta_1 \approx 31$meV. As the interactions between electron pockets are likely  smaller than between electron and hole pockets,
we set, for definiteness, $U_5=U_2=0.5U_1$, $U_1=U_3$,  and $U_{j1}=0.1U_j$.
We verified that the spin wave dispersion around the ordering wave vector
does not change in any substantial way if we vary these numbers. In particular, when the ellipticity parameter $\epsilon \neq 0$, we can set $U_{j1} =0$ and it will not affect the outcome.
\begin{figure}[t]
\centering
\includegraphics[width=1.0\linewidth]{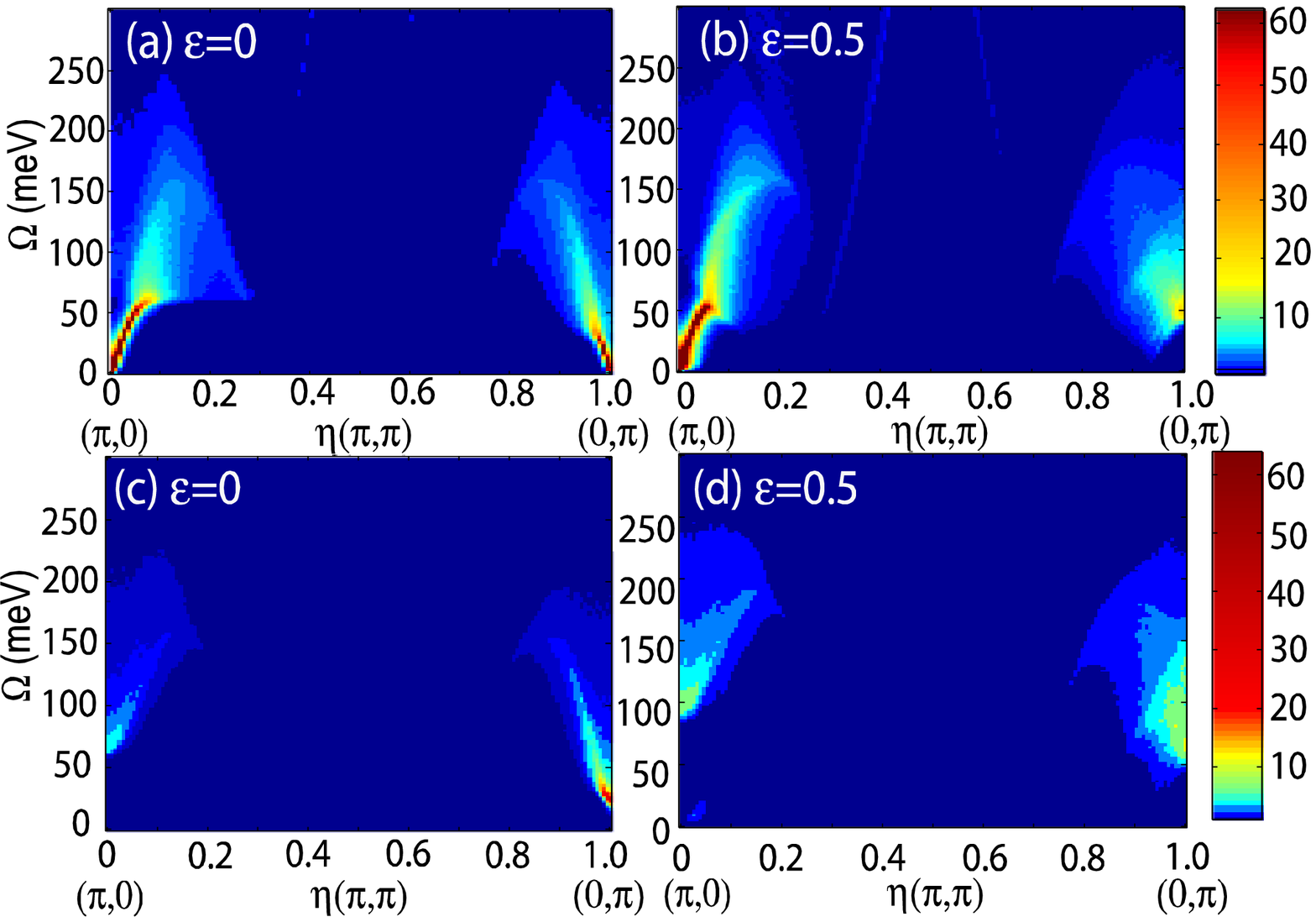}
\caption{(color online) Calculated imaginary part of the transverse (upper panel) and longitudinal (lower panel) component of the physical RPA spin susceptibility as a function
of frequency and momentum (plotted along $(0,\pi) \to (\pi,0)$ points of the BZ) for
$\epsilon=0$ (a)-(c), and $\epsilon=0.5$ (b)-(d). The color bars refer to the intensity in
units of st./eV.  For numerical purposes we set the damping constant $\delta = 1/300$eV.}
\label{fig3}
\end{figure}
The results of our calculations are presented in Fig. \ref{fig3} for $T=0$ and
for several values of the ellipticity parameter $\epsilon$.
Consider first the case of circular electron pockets, $\epsilon =0$, which is for magnetic properties
the same as full nesting. In this case two out of three FS are completely gapped, if the order parameter exceeds the threshold value, which is the case for our $\Delta$.
In Fig.\ref{fig3}(a) we show the calculated
imaginary part of the transverse component of the susceptibility, Im
$\left[\hat\chi_{+-} ({\bf q},\Omega)\right]_{RPA}$ along the direction
$(0,\pi)$ to $(\pi,0)$. Because the corresponding ground state is degenerate, the excitation spectrum  has a Goldstone mode at the ordering momentum ${\bf Q}_1$ and another
gapless mode at ${\bf Q}_2$.
This result was earlier found in Ref.~\cite{brydon_new}. We also clearly see that the excitations near
the ordering momentum are propagating up to $\Omega_c({\bf Q}_1) = 2\Delta_1 \sim 60 meV$ what is expected because both pockets separated by ${\bf Q}_1$ are gapped. Excitations near ${\bf Q}_2$ are also propagating, but up to smaller energies,  $\Omega_c({\bf Q}_2) =\Delta_1$, that is consistent with the fact that only one of the two pockets separated by ${\bf Q}_2$ is gapped. At higher energies spin-waves  enter the continuum
and become overdamped Stoner-like excitations, however, still with well defined peaks.
The difference between $\Omega_c({\bf Q}_1)$ and $\Omega_c({\bf Q}_2)$ is also seen in Fig.\ref{fig3}(c) where we plotted
the longitudinal component of the susceptibility. As expected,
 Im$\left[\hat\chi_{+-} ({\bf q},\Omega)\right]_{RPA}$ vanishes below  $\Omega_c({\bf Q})$.

At finite $\epsilon =0.5$, we used a slightly larger $U_{SDW}$ to recover the same $T_N$ as for $\epsilon=0$, which in turn leads to a slightly larger $\Delta_1$. The results are presented in panels (b) and (d) of Fig. \ref{fig3}. There are two key effects introduced by ellipticity.
 First,  the degeneracy is now broken and the transverse excitations near ${\bf Q}_2$ acquire a finite gap clearly visible in Fig.\ref{fig3}(c). Second,
 spin-wave excitations near ${\bf Q}_1$ have a finite Landau damping down to
$\Omega =0$ because the SDW order no longer completely gaps the hole and electron FSs separated by ${\bf Q}_1$.
At the same time, spin-wave excitations are still clearly visible at low energies and do not become overdamped. The reasoning is that SDW coherence factors
 suppress the scattering between $\alpha$ and $\beta_1$ fermions and
reduce the Landau damping around the ordering momentum ${\bf Q}_1$  from $\Omega/|{\bf q}-{\bf Q}_1|$ to $\Omega |{\bf q}-{\bf Q}_1|$ what for linear dispersion is of the same order as $\Omega^2$ and $({\bf q}-{\bf Q}_1)^2$.
 Within the one-band model, this effect has been discussed in~\cite{chubukov_last}.
Once $\Omega$ exceeds $\Omega_c ({\bf Q}_2)$, the coherence factors no longer screen Landau damping, and spin-waves  become Stoner-like, overdamped excitations.
 The scattering between $c$ and $\beta_2$ fermions is not
suppressed by the SDW coherence factors because ${\bf Q}_2$ is not the ordering momentum, therefore, Landau damping makes the transverse excitations near ${\bf Q}_2$ overdamped immediately above the gap. Note, that there is also a tendency towards incommensuration near ${\bf Q}_2$.
 Longitudinal excitations are still gapped up to $\Omega_c ({\bf Q})$ and become Stoner-like at higher energies, as is clearly seen in Fig. \ref{fig3}(d).

\begin{figure}[t]
\includegraphics[width=1.0\linewidth]{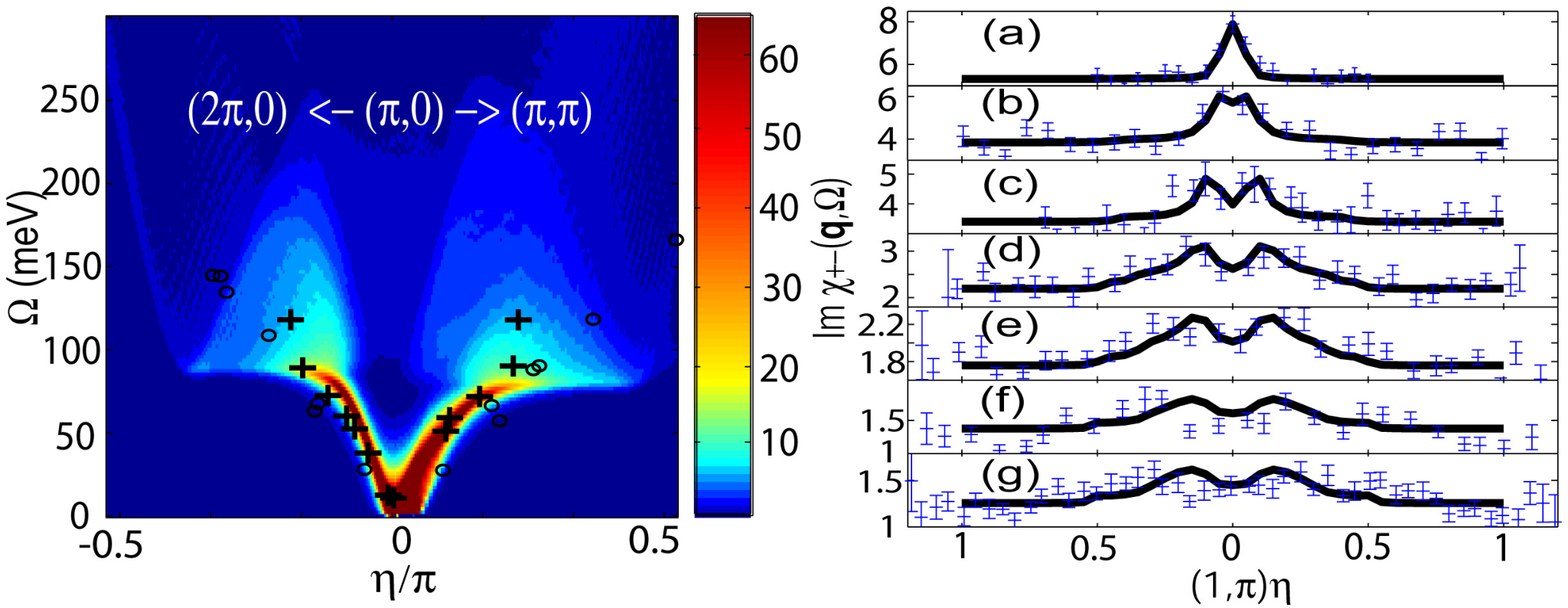}
\caption{(color online) Left panel: Calculated Im$\chi^{+-}_{RPA}$ as a function of frequency and momentum (plotted along $(0,2\pi) \to (0,\pi) \to
(\pi,\pi)$ directions of the BZ) for  $\epsilon=0.5$. The crosses and circles are the measured points
taken from Refs. \cite{Diallo} and \cite{zhao}, respectively. Right panel: Calculated constant energy cuts of Im$\chi^{+-} ({\bf q}, \Omega)$ along $(0,0) \to (\pi,0) \to (2\pi, 0)$
direction (solid curves) for 25 (a), 48 (b), 64 (c), 100 (d), 115 (e),  135 (f), and 144 (g) (all in meV). The experimental
data are taken from Ref.\cite{zhao}. The intensity maxima have been adjusted to the experimental values.}
\label{fig4}
\end{figure}
Despite the fact that our model is indeed a simplification of the actual 5-band model for FPs, it reproduces quite well  the experimental
data for the magnetic excitations. First, the excitations near the ``wrong'' momentum ${\bf Q}_2$ are gapped up to about $~$50meV as seen in Fig.\ref{fig3}(c). Second, measured excitations near ${\bf Q}_1$ are propagating spin-waves
up to about $100$ meV, with a finite but small damping. In the left panel of
Fig.\ref{fig4} we compare our dispersion with  the experimental data~
\cite{zhao,Diallo}. We emphasize that both the measured and the calculated spectra
have different velocities along different crystallographic directions. In our theory, it is a consequence of the non-zero ellipticity {\it and} the fact that only one electron pocket is involved in the SDW formation.
Remarkably, for $\epsilon=0.5$ the anisotropy is the same as in the experimental data. Another verifiable
theory prediction for  this range is that the width of the spin-wave peak
  should scale as $\Omega^2$. Third, we find that excitations are still visible even  when the spin-waves enter the continuum. In the right panel of Fig.\ref{fig4}  we compare the calculated and the measured $Im \chi^{+-} (q, \Omega)$ along  $(0,0) \to (\pi,0) \to (2\pi, 0)$ direction for different $\Omega$. We see that the agreement is quite good for all frequencies even above $\Omega_c$ \cite{final_remark}.

{\it Conclusions}
In this paper we analyzed the structure of the magnetic excitations in
 the magnetically ordered state of Fe-pnictides. We used a multi-band
itinerant model and developed a multi-band RPA treatment of the longitudinal and
transverse components of the dynamical spin susceptibility.
 We found weakly-damped  spin-waves near the ordering momentum and showed how they dissolve into the particle-hole continuum above an energy which scales with the SDW order parameter. For perfect nesting between
 electron and hole bands the SDW state has an extra degeneracy, and we found
 an extra  gapless mode at the momentum different from the ordering one.
 When  ellipticity of the electron bands is included, the  degeneracy is
lifted and the extra mode becomes gapped. Ellipticity of the electron pockets also naturally accounts for the anisotropy of the spin waves along different crystallographic directions.  We argue that our theory agrees well with the available neutron scattering data.

We thank P. Dai and J. Zhao for sending us the experimental data and N. Shannon, M. Vavilov and A. Vorontsov
 for stimulating discussions. I.E. acknowledges
the support from the RMES Program (Contract No. N
2.1.1/3199), and NSF (Grant DMR-0456669).  A.V.C. acknowledges the support from NSF-DMR 0906953.

\end{document}